\documentclass[twocolumn,numbering,showpacs,superscriptaddress]{revtex4-1}

\usepackage{graphicx,color}
\usepackage{latexsym}
\usepackage{amsmath,amssymb}        
\usepackage[draft=false]{hyperref}
\usepackage{mathrsfs}
\usepackage{comment}
\usepackage{soul}
\usepackage{mathrsfs}
\definecolor{orange}{rgb}{1,0.5,0}
\DeclareSymbolFontAlphabet{\mathrsfs}{rsfs}
\DeclareMathAlphabet{\mathcal}{OMS}{cmsy}{m}{n}

\usepackage{ulem}

\begin{document}

\title{The CMB and MPS profiles for a novel and efficient model of dark energy}

\author{Tonatiuh Matos}
\email{tonatiuh.matos@cinvestav.mx}
\affiliation{Departamento de F\'{i}sica, Centro de Investigaci\'on y de Estudios Avanzados del IPN, A.P. 14-740, 07000 CDMX, M\'exico.}

\author{Luis Osvaldo Tellez-Tovar}
\email{ltellez@fis.cinvestav.mx}
\affiliation{Departamento de F\'{i}sica, Centro de Investigaci\'on y de Estudios Avanzados del IPN, A.P. 14-740, 07000 CDMX, M\'exico.}
\affiliation{Instituto de Ciencias F\'isicas, Universidad Nacional Aut\'onoma de M\'exico, 
Apdo. Postal 48-3, 62251 Cuernavaca, Morelos, M\'exico.}

\begin{abstract}
In a previous work \cite{Laura} it was shown that by considering the quantum nature of the gravitational field mediator, it is possible to introduce the momentum energy of the graviton into the Einstein equations as an effective cosmological constant. The Compton Mass Dark Energy (CMaDE) model proposes that this momentum can be interpreted as dark energy, with a
Compton wavelength given by the size of the observable universe $R_H$, implying that the dark energy varies depending on this size. The main result of this previous work is the existence of an effective cosmological constant $\Lambda=2\pi^2/\lambda^2$ that varies very slowly, being $\lambda=(c/H_0) R_H$ the graviton Compton wavelength. In the present work we use that the dark energy density parameter is given by $\Omega_\Lambda=2\pi^2/3/R_H^2$, it only has the curvature $\Omega_k$ as a free constant and depends exclusively on the radiation density parameter $\Omega_r$. Using $\Omega_{0r} = 9.54\times10^{- 5}$, the theoretical prediction  for a flat universe of the dark energy density parameter is $\Omega_{0\Lambda} = 0.6922$. We perform a general study for a non-flat universe, using the Planck data and a modified version of the CLASS code we find an excellent concordance with the Cosmic Microwave Background and Mass Power Spectrum profiles, provided that the Hubble parameter today is $H_0 = 72.6$ km/s/Mpc for an universe with curvature $\Omega_{0k}=-0.003$. We conclude that the CMaDE model provides a natural explanation for the accelerated expansion and the coincidence problem of the universe.

\end{abstract}

\pacs{Cosmological Constant -- Hubble Parameter-- Compton Mass}

\maketitle

Without a doubt, one of the most important problems facing science today is that of explaining the accelerating expansion of the universe. Since 1998, with observations of SNIa-type supernovae it was established that the universe is experiencing a clear accelerated expansion contrary to the belief that the expansion must be slowing down due to the gravitational force of all matter in the universe itself. Since that time several independent tests have been conducted for the same observation, today there is no doubt that the universe is accelerating. The question has provoked an enormous amount of hypotheses and explanations, from the simple cosmological constant, proposed by Einstein himself to the modification of Einstein's equations, massive gravity, hollographic universe, etc.

One of the beliefs is that the explanation for the acceleration of the universe could come from quantum mechanics, that is, from a theory of quantum gravity. This possibility is robust and has been explored by various scientific groups around the world, sadly without success. In the reference \cite{Laura} they proceeded in an alternative way, because up to now we do not have a theory of quantum gravity, in this reference the authors propose an effective way to introduce the quantum character of the graviton, using analogies with other fields and interactions. They show that with this proposal the system behaves very similar to the $\Lambda$CDM case. The similarity was excellent and this hypothesis led to further studies. In this work we will show that the predictions of the $\Lambda$CDM and CMaDE models are indistinguishable, at least at cosmological scales, since the CMB and MPS profiles, the two strongest observations we have in the universe, are exactly the same, but the CMaDE model using an explanation of quantum mechanics without dark energy.

In this work we want to study a possible solution given in \cite{Laura} for the last problem using very simple arguments for the gravitational interaction. The main goal of this work is not to convince the reader of the arguments given in \cite{Laura} to find a form for dark energy, but to use this form as an effective function, a proposal to fit all the observations without free constants. In what follows we remain the main ideas of \cite{Laura}, but then we use the functional form of the dark energy in effective way.

The main arguments of \cite{Laura} is that in the case of a massless particle, such as the gravitational interaction mediator or graviton, the energy due to its momentum $E=pc$, is not contained in the Einstein equations. In the Einstein's equations it is implicit that the mass of the mediator of the gravitational interaction is zero. On the other side, the energy of the graviton due to its moment comes from the quantum mechanical character of the graviton. But everything in nature gravitates.
The claim of \cite{Laura} is that this energy also gravitates and must be counted as extra energy. 

The hypotheses in \cite{Laura} are: if Gravitation is a quantum mechanical interaction its mediator has a Compton effective mass and its corresponding wavelength $\lambda$ is limited by the size of the observable universe. Using these arguments, they found that the cosmological constant is given by
\begin{equation}\label{eq:Lam}
     \Lambda=\frac{2\pi^2}{\lambda^2}.
 \end{equation}
We will use $\Lambda$ as indicated in (\ref{eq:Lam}) as an effective result, where $\Lambda$ varies very slowly, as we shall see.

For an observer today, the gravitational interaction travels a distance $R_H$ during its life, the wavelength will be $\lambda=(c/H_0) R_H$ long, where $R_H$ is the unitless length

\begin{equation}\label{eq:RH}
 R_H=H_0\int_0^{t}\frac{dt'}{a}=\int_{-\infty}^N\frac{H_0}{H}e^{-N'}dN',
 \end{equation}
given in terms of the e-folding parameter $N=\ln(a/a_0)$ and the Hubble parameter $H=\dot N$, being $a$ the scale factor of the universe and $a_0$ its value today. 
\begin{figure}
\centering
\includegraphics[width=8.cm]{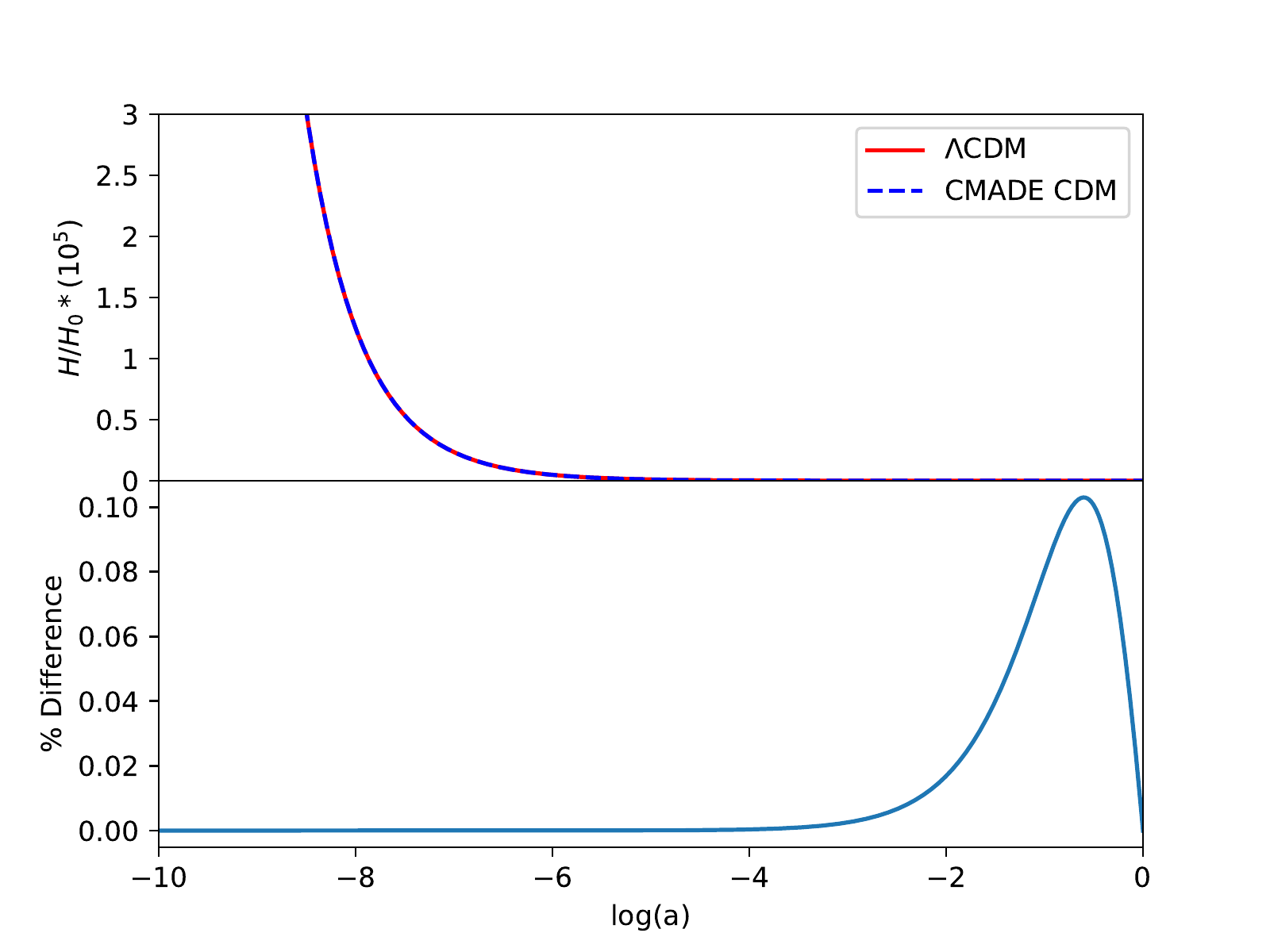}
\caption{In the upper panel we show the evolution of the Hubble parameter using the CMaDE (equation (\ref{eq:Friedmann2}), point line) and the $\Lambda$CDM model (solid line). We used the Planck values $\Omega_{0m}=0.315$, $\Omega_{0r}=10^{-4}$, $H_0=67.3$ km/s/Mpc in both plots and the value $\Omega_{0\Lambda}=0.684$ for the $\Lambda$CDM curve and (\ref{eq:Lam2}) for the CMaDE model. In the lower panel we show the proportional difference between both curves $(H_{CMaDE}-H_{\Lambda CDM})/H_{\Lambda CDM}$.}
\label{fig:HNumerico}
\end{figure}

In order to obtain the Friedmann equation for our model, we consider that
\begin{equation}\label{eq:Friedmann}
     H^2+\frac{k}{a^2}=\frac{\kappa^2}{3}\left(\rho_m+\rho_r+\rho_\Lambda\right),
 \end{equation}
here $\rho_m$ represents the matter density, $\rho_r$ corresponds to radiation density, $k$ is the curvature parameter and $\rho_\Lambda=\Lambda c^2 /\kappa^2$ is the dark energy density, where $\kappa=8\pi G/c^4$ is the Einstein constant. Using equations (\ref{eq:Lam}) and (\ref{eq:RH}) in the derivative of (\ref{eq:Friedmann}), we get that \cite{Laura}
\begin{eqnarray}\label{eq:Friedmann2}
      \frac{HH'}{H_0^2}+\Omega_{0k} e^{-2N}+\frac{3}{2}\Omega_{0m} e^{-3N}+2\Omega_{0r} e^{-4N}&-&\nonumber\\\sqrt{\frac{3}{2}}\frac{H_0e^{-N}}{H\pi}\left(\frac{H^2}{H_0^2}-\Omega_{0k} e^{-2N}-\Omega_{0m} e^{-3N}-\Omega_{0r} e^{-4N}\right)^{\frac{3}{2}}&=&0,\nonumber\\
\end{eqnarray}
where, for any given variable $q$, the prime means $q' = dq/dN = \dot{q}/H$ and $\Omega_{0x}=\rho_{0x}/\rho_{crit}$, with the critical density of the universe today given by $\rho_{crit}=3H_0^2/\kappa^2$. In particular
\begin{equation}\label{eq:OL}
    \Omega_\Lambda=\frac{2}{3}\frac{\pi^2}{R_H^2}.
\end{equation}
From the above we have that the CMaDE Friedmann equation is given by (\ref{eq:Friedmann2}).

Here it is important to note that given (\ref{eq:Lam}) with (\ref{eq:RH}) for the function $\Lambda$ implies that the CMaDE model only has curvature as a free constant to fit all observations. If we integrate (\ref{eq:Friedmann2}) and (\ref{eq:RH}) we find that $R_H\sim 3$. Also note that, because $\Lambda$ is not a constant, the Bianchi identities have an extra term 
\begin{eqnarray}\label{eq:Ldot}
      \dot\Lambda=H\frac{d\Lambda}{dN}=-4\pi^2\left(\frac{H_0}{c}\right)^2H_0\frac{e^{-N}}{R_H^3}.
\end{eqnarray}
We obtain that $\dot\Lambda=-4.48\times 10^{-16}h_0^3/R_H^3$/Mpc$^{2}$/yr; its value today is $\dot\Lambda\sim- 5\times 10^{-17
}h_0^3$/Mpc$^{2}$/yr, which is really very small, being $H_0=100h_0$Km/sec/Mpc. Note that just after inflation we can put that $\dot\Lambda = -1.5\times 10^{-72}h_0^3e^{-N}/R_H^3$/cm$^2$/sec, which depends on the value of $R_H$. However, the redshift for inflation is $z\sim 10^{26}$, this means that $N\sim -60$. So, before inflation the wavelength is small, the exponential factor is big and the Bianchi identities have an extra term given by (\ref{eq:Ldot}). 

When inflation ends, the wavelength grows up about $e^{60}$ times thus $R_H\sim \lambda_0e^{60}$ grows very fast and (\ref{eq:Ldot}) becomes very small. This means that the Bianchi identities hold up very well, because $\dot\Lambda\sim 0$, i.e., after inflation $\Lambda$ is almost constant. So the equation (\ref{eq:Lam}) can be viewed as a very slowly varying cosmological constant.

So far, the arguments used in \cite{Laura} could be controversial for some readers, the objective of this work is not to discuss these arguments, but to see equation (\ref{eq:Lam}) with the integral (\ref{eq:RH}) as an effective proposal and check if them can explain the observable universe, leaving for future work the possible quantum gravity explanation of the equations (\ref{eq:Lam}) and (\ref{eq:RH}) \cite{Omar}. Note that this $\lambda$ is similar to the proposal of holographic dark energy where we know that this model is not able to explain the dark energy behavior of the universe \cite{delCampo:2011jp}. The difference of (\ref{eq:RH}) with the holographic model proposal is that the holographic wavelength is the distance to the horizon of the universe, this integral has an extra scale factor outside the corresponding integral (\ref{eq:RH}). The other main difference is that the holographic model has a free constant in the cosmological function $\Lambda$, while the equation (\ref{eq:Lam}) has no free parameters. So, let us think of the equation (\ref{eq:Lam}) as an effective proposal and its justification are the results that we find in this work.

In fig.\ref{fig:HNumerico} we compare the numerical solution of (\ref{eq:Friedmann2}) with the evolution of the Hubble parameter in $\Lambda$CDM, $H_{\Lambda CDM}=H_0\sqrt{\Omega_{0m} e^{-3N}+\Omega_{0r} e^{-4N}+\Omega_\Lambda}$. Note that $H$ has the same evolution for both models implying same predictions. Note too that the CMaDE density remains subdominant for large redshifts and is a bit different than the evolution of LCDM for small redshifts. 

Solving numerically (\ref{eq:Friedmann2}) for a flat space-time we carry out the integral (\ref{eq:RH}) and we find that $R_H = 3.083$ in (\ref{eq:Lam}). 
With these results we obtain that
\begin{equation}\label{eq:Lam2}
    \Lambda=2\left(\frac{\pi}{3.087}\right)^2\frac{H_0^2}{c^2}=\frac{3H_0^2}{c^2}\Omega_{0\Lambda},
 \end{equation}
We can see that the value of $\Omega_{\Lambda}$ strongly depends on the size of the wavelength. (\ref{eq:RH}).

We can use the size of the universe horizon to determine the value of the wavelength $\lambda$. Thus we can determine the value of the CMaDE now and give an explanation of the cosmological and  coincidence problems. 

\begin{figure}
\centering
\includegraphics[width=6.0cm]{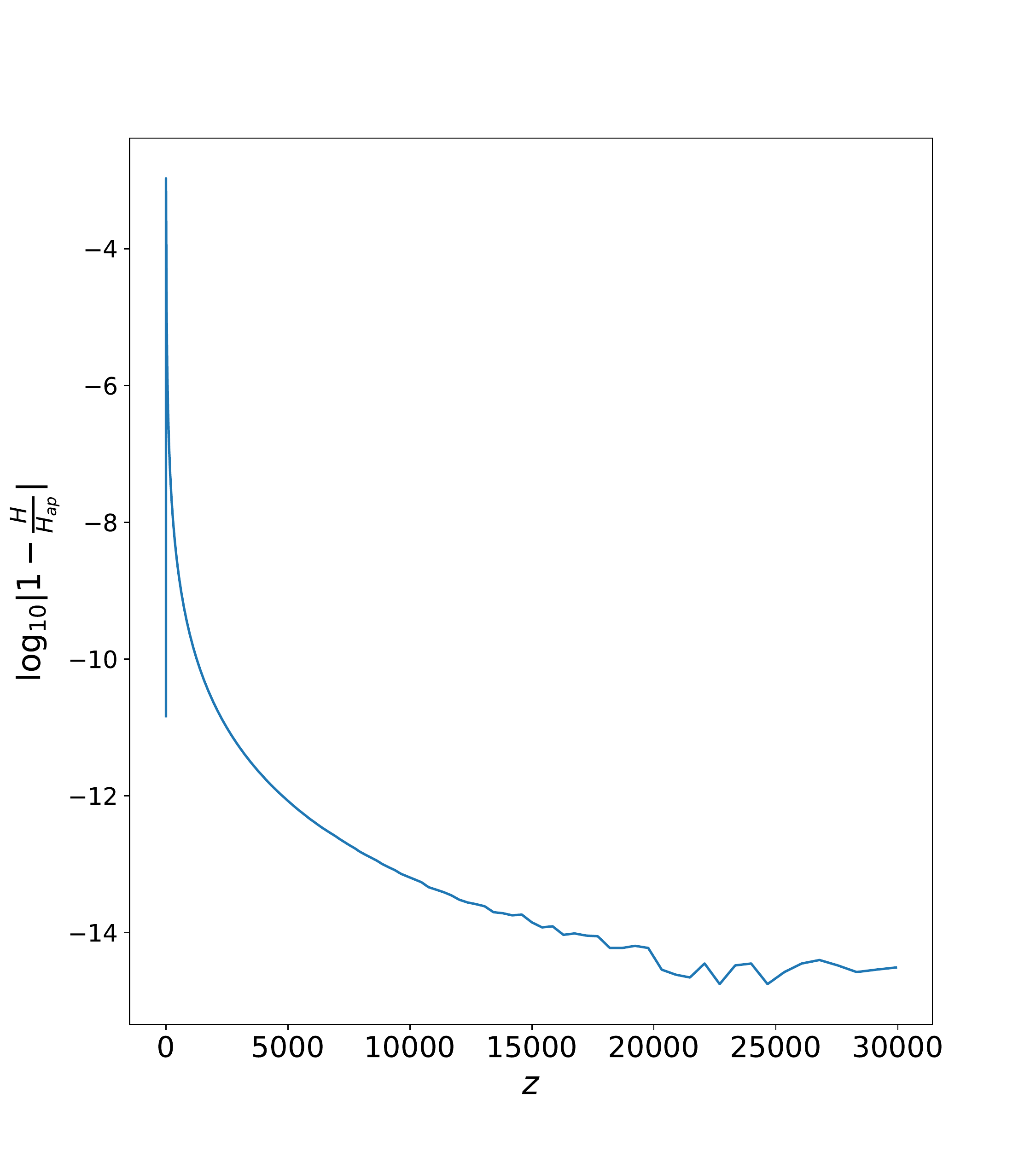}
\caption{Evolution of the $1-H/H_{app}$ using the numerical integration of (\ref{eq:Friedmann2}) ($H$) and equation (\ref{eq:H0RD}) ($H_{ap}$), with $q=0.695$. We plot $\log(|1-H/H_{app}|)$, observe that this ratio is always less than $10^{-3}$. }
\label{fig:HvsH}
\end{figure}

\begin{figure}
\centering
\includegraphics[width=8.0cm]{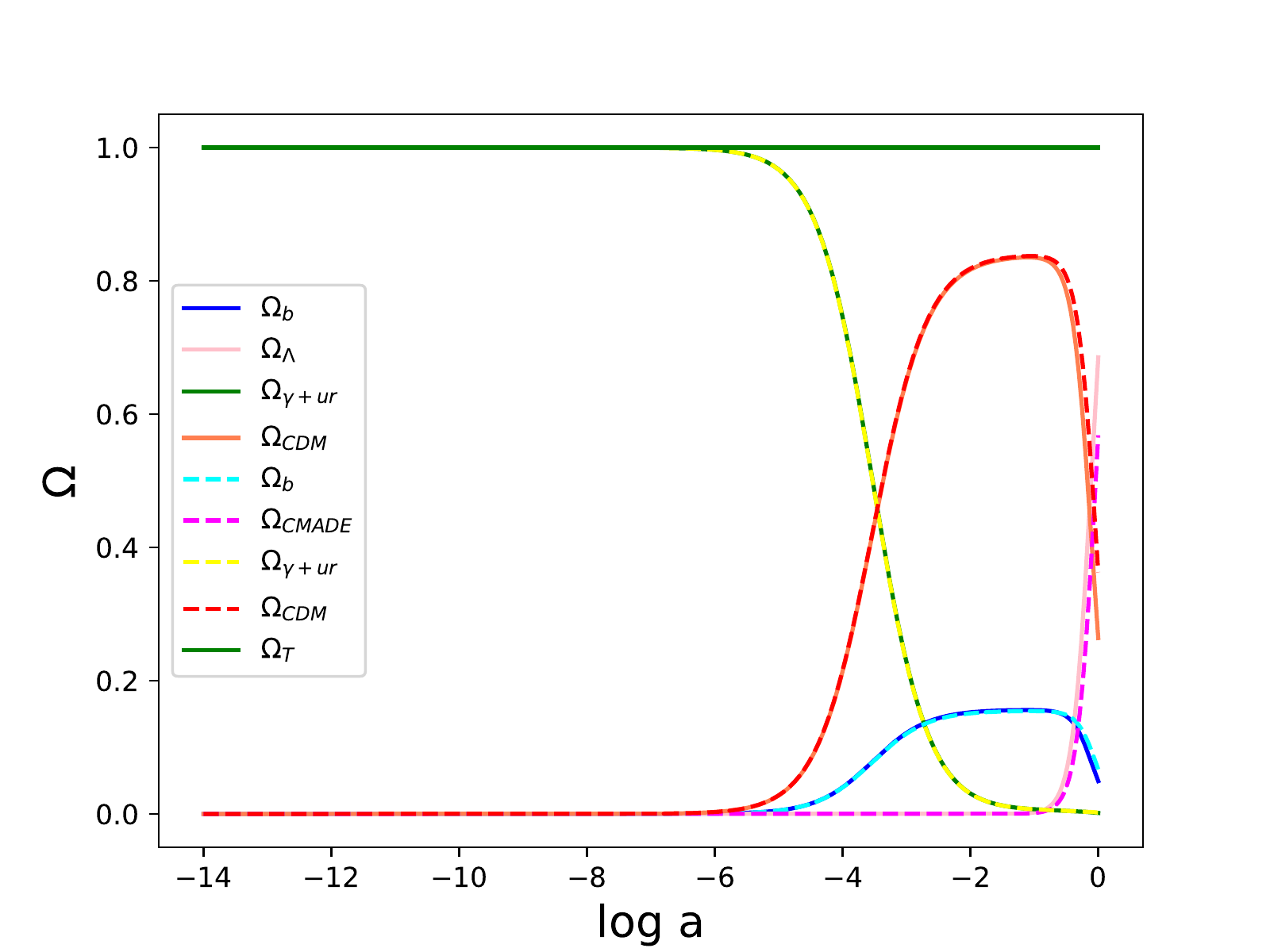}
\caption{Evolution of the $\Omega$'s using equations (\ref{eq:H0MD}) and (\ref{eq:H0RD}) and the corresponding ones using the $\Lambda$CDM model.}
\label{fig:Omegas}
\end{figure}

\begin{figure}
\centering
\includegraphics[width=0.45\textwidth]{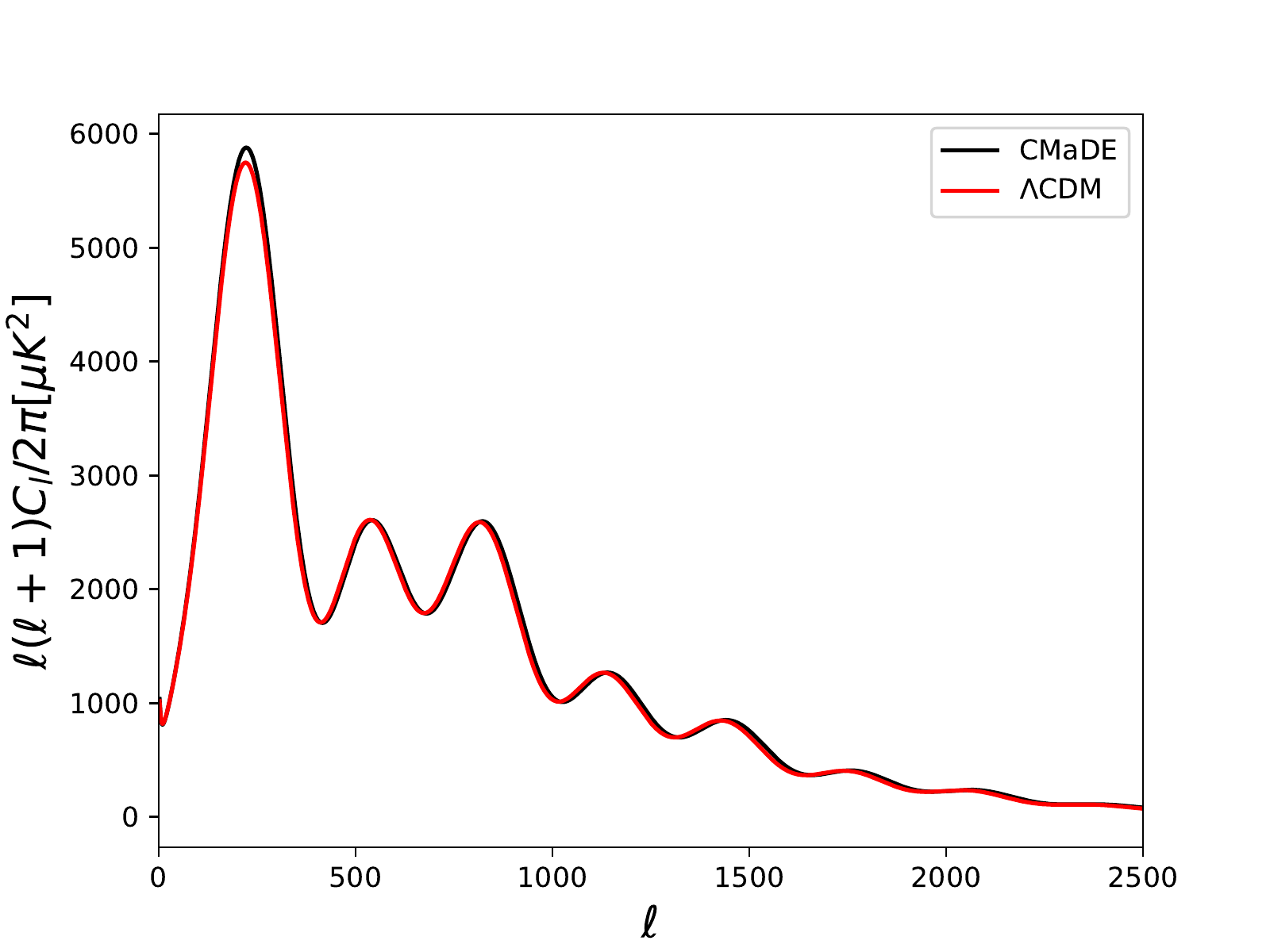}
\includegraphics[width=0.45\textwidth]{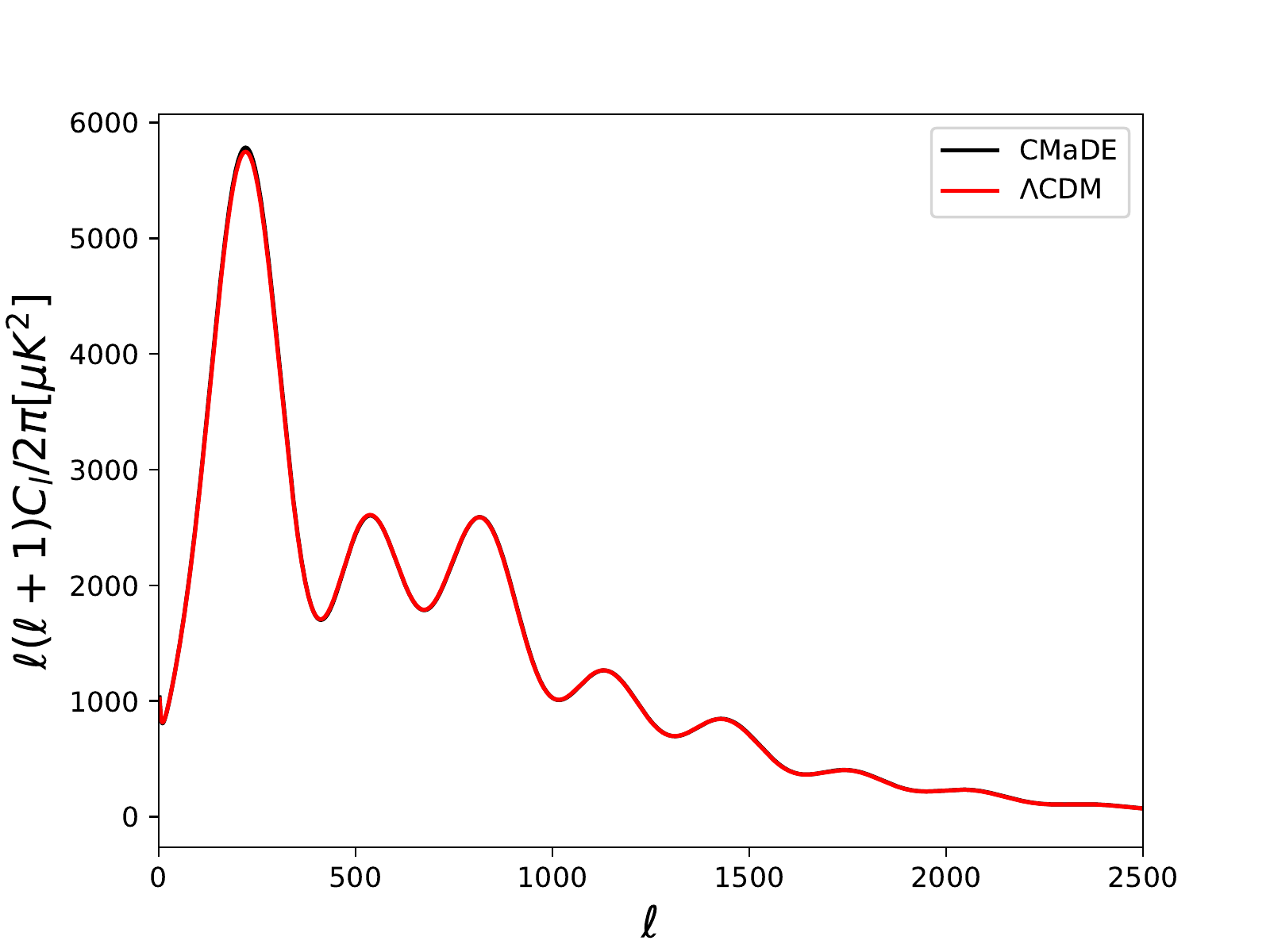}
\includegraphics[width=0.45\textwidth]{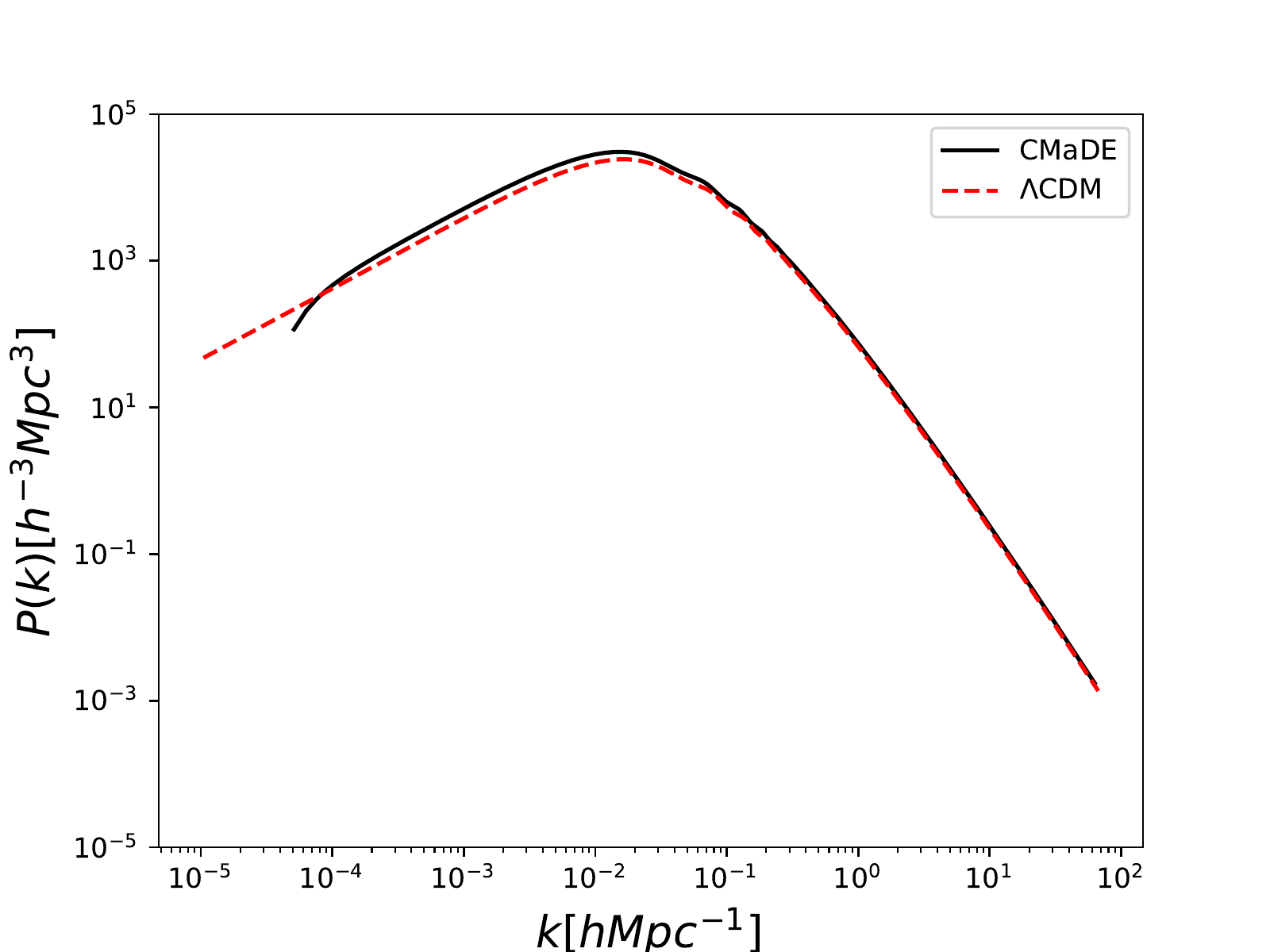}
\caption{Profiles of the CMB for a flat universe (upper panel) and for a closed universe with $\Omega_{0k}=-0.003$ (middle panel) and MPS (lower panel) observations using an amended version of CLASS code \cite{2011JCAP...07..034B}. We compare them with the best fit of the $\Lambda$CDM model, using data from the Planck satellite. Note that the CMB temperature fluctuations for the flat universe are the same as the $\Lambda$CDM, the only difference is in the first maximum. For the MPS there are very small discrepancies for the small structure. The CMaDE model settings are  $\Omega_{0r}=5.67\times10^{-5}$, $q=0.694$, $H_0=72.6$ km/s/Mpc and $\Omega_{0b}=0.044$ for the flat universe and $q=0.695$, $\Omega_{0k}=-0.003$, $H_0=72.6$ km/s/Mpc and $\Omega_{0b}=0.043$ for the closed universe. Observe that the value of $H_0$ is very close to the observed one from the local distance ladder \cite{Riess:2016jrr}}
\label{fig:Hevolution}
\end{figure}

In what follows we want to study the possibility that the CMaDE model is capable of reproducing all the observations of the universe that we have so far. Strictly speaking we have to solve equation (\ref{eq:Friedmann2}) and solve the whole cosmology using it \cite{Laura}\cite{Osvaldo}. However, in this work we first solve the entire cosmology using an approximation. Here we will focus on the temperature fluctuations of the cosmic microwave background (CMB) and the mass power spectrum (MPS) only, leaving a more in-depth analysis of the rest for future work. \cite{Osvaldo}. In order to find a suitable approximation to the equation (\ref{eq:Friedmann2}), we proceed as follows. We know that during the epoch dominated by matter $H = 2/(3t) = H_0/a^{3/2}$ \cite {Weinberg2}, so we found that the evolution of $R_H$ is given by $R_H = 2\sqrt{a}$. 
Thus, during this time we have that 
\begin{equation}\label{eq:Lam3}
     \Lambda^{MD}\sim\frac{\pi^2}{6}\frac{3H_0^2}{c^2}\frac{1}{4a}.
 \end{equation}

So, we find that the field equation for $\Lambda^{MD}$ is $\dot\Lambda^{MD}+H\Lambda^{MD}=0$. We use this approximation to get the Hubble parameter evolution, given as follows

\begin{equation}\label{eq:H0MD}
     H=H_0\sqrt{\Omega_{0m} e^{-3N}+\Omega_{0r} e^{-4N}+\Omega_{0k} e^{-2N}+\Omega_{0\Lambda} e^{-N}}.
 \end{equation}
 However, this approximation is not good enough for the numerical solution of (\ref{eq:Friedmann2}). Instead of that we will approximate it with the function
 
 \begin{equation}\label{eq:H0RD}
     H_{app}=H_0\sqrt{\Omega_{0m} e^{-3N}+\Omega_{0r} e^{-4N}+\Omega_{0k} e^{-2N}+\Omega_{0\Lambda}  e^{q N}}.
 \end{equation}
 where $q$ and $\Omega_{0\Lambda}$ are constants that fit the numerical solution. The similarity between the function (\ref{eq:H0RD}) with the numerical integration of (\ref{eq:Friedmann2}) is very good everywhere, see figure \ref{fig:HvsH}.
 
 The radiation content of the universe, CMB photons plus neutrinos, is given by $\rho_r=2(1+3\times 7/8(4/11)^{4/3})T^4$. The CMB observations indicate that $T=2.7255$ K, thus  $\Omega_{0r}=9.54\times10^{-5}$. We set $\Omega_{0\Lambda}$ such that $H_{app}=H_0$ at $N=0$. These values, again, are very close to that obtained in $\Lambda$CDM. 
 
 In fig. \ref{fig:Omegas} we see the evolution of the $\Omega$'s for the CMaDE model, using the function (\ref{eq:H0RD}) and the $\Lambda$CDM model, we see the similarity of the evolution.
 
 Thus, the next step is to see whether this approximation gives us the correct behavior of the CMB and MPS profiles. In figure \ref{fig:Hevolution} we see the comparison between the profiles of the CMaDE and $\Lambda$CDM models using an amended version of CLASS code \cite{2011JCAP...07..034B}, again the similarity between both models is very close. The only difference we find for the flat universe is an excess of temperature predicted by the CMaDE model in the first maximum, but in the rest, of the two profiles, the coincidence with the observations using the Planck data is very good. It is very remarkable that the value of $\Omega_\Lambda$ in the CMaDE model is completely theoretical, so it is quite relevant that this match with the observations is so good. We believe that the small differences could be due to the fact that we are using an approximation for the CMaDE model and not the solution of the equation (\ref{eq:Friedmann2}) or by some extra phenomenon. However, in this work we want to present the main characteristics of the CMaDE model, the observational aspects of the model will be found elsewhere \cite{Osvaldo}. 

Finally, considering the gravitational field quantum nature we found that if it has a quantum Compton effective mass we could see it as a variable  ``cosmological constant". With these result we could explain the actual value of the density parameter of the dark energy and the coincidence problem. Nevertheless, we think that this hypothesis opens a new window of research and must be further studied.

\acknowledgments

We thank Alejandro Avilés for the approximation (\ref{eq:H0RD}), Jorge Cervantes and Mario Rodriguez for helpful discussions. This work was partially supported by CONACyT M\'exico under grants  A1-S-8742, 304001, 376127, 240512;
The authors are gratefully for the computing time granted by LANCAD and CONACYT in the
Supercomputer Hybrid Cluster "Xiuhcoatl" at GENERAL COORDINATION OF INFORMATION AND
COMMUNICATIONS TECHNOLOGIES (CGSTIC) of CINVESTAV.
URL: https://clusterhibrido.cinvestav.mx/ and Abacus clusters at Cinvestav, IPN;
I0101/131/07 C-234/07 of the Instituto
Avanzado de Cosmolog\'ia (IAC) collaboration (http://www.iac.edu.mx/).

\bibliography{main}{}

\end{document}